# Fundamental Solutions for Micropolar Fluids

JIAN-JUN SHU[*] and JENN SHIUN LEE

School of Mechanical & Aerospace Engineering,

Nanyang Technological University, 50 Nanyang Avenue, Singapore 639798

## Abstract

New fundamental solutions for micropolar fluids are derived in explicit form for two- and three-dimensional steady unbounded Stokes and Oseen flows due to a point force and a point couple, including the two-dimensional micropolar Stokeslet, the two- and three-dimensional micropolar Stokes couplet, the three-dimensional micropolar Oseenlet, and the three-dimensional micropolar Oseen couplet. These fundamental solutions do not exist in Newtonian flow due to the absence of microrotation velocity field. The flow due to these singularities is useful for understanding and studying microscale flows. As an application, the drag coefficients for a solid sphere or a circular cylinder that translates in a low-Reynolds-number micropolar flow are determined and compared with those corresponding to Newtonian flow. The drag coefficients in a micropolar fluid are greater than those in a Newtonian fluid.

## 1. Introduction

The physical mechanisms of heat, mass, and momentum transport in small-scale units may differ significantly from those in macroscale equipment [1,2]. Fundamental and applied investigations of microscale phenomena in fluid mechanics are motivated by developments in the areas of biological molecular machinery, atherogenesis, microcirculation, and microfluidics. At scales larger than a micron, the fluid can be treated as a continuum, and the flow is governed by the Navier-Stokes equation. The continuum model assumes that the properties of the material vary continuously throughout the flow domain. In Newtonian continuum mechanics, the fluid is modeled as a dense aggregate of particles, possessing mass, and translational velocity. However, the field equation, such as the Navier-Stokes equation, does not account for the rotational effects of the fluid micro-constituents.

---

[*] Author to whom correspondence should be addressed.

In the theory of micropolar fluids [3], rigid particles contained in a small volume element can rotate about the centroid of the volume element. The rotation is described by an independent micro-rotation vector. Micropolar fluids can support body couples and exhibit microrotational effects. The theory of micropolar fluids has shown promise for predicting fluid behaviour at microscale. Papautsky, *et al.* [1] found that a numerical model for water flow in microchannels based on theory of micropolar fluids gave better predictions of experimental results than those obtained using the Navier-Stokes equation. Micropolar fluids can model anisotropic fluids, liquid crystals with rigid molecules, magnetic fluids, clouds with dust, muddy fluids, and some biological fluids [3]. In view of their potential application in microscale fluid mechanics and non-Newtonian fluid mechanics, it is worth exploring new fundamental solutions.

The fundamental solutions for Stokes flow [4] and Oseen flow [5] due to a point force are commonly named as the Stokeslet and the Oseenlet. The fundamental solution due to a point force in a steady Stokes flow was first derived by the Nobel Laureate, Lorentz, as far back as 1896 [6]. This solution is now known by the name Stokeslet, although Stokes never knew about it. The name Stokeslet was coined by Hancock in 1953 [7]. Obviously, Hancock was unaware of this Lorentz's work. Had he known about it, he might have opted for Lorentzlet instead of Stokeslet. There are many who believe that the steady Hancock-named Stokeslet was derived by Ladyzhenskaya in 1961 [8] by using Fourier-transform methods. A number of more appropriate references concerning the steady Stokeslet and Oseenlet can be found in [9]. The closed-form fundamental solutions for generalized unsteady Stokes and Oseen flows associated with arbitrary time-dependent translational and rotational motions have been derived by Shu & Chwang in 2001 [10].

In micropolar fluids, the microrotation fundamental solutions due to a point force are the micropolar Stokeslet and micropolar Oseenlet, and those due to a point couple are the micropolar Stokes couplet and micropolar Oseen couplet. Such fundamental solutions do not exist in Newtonian flow due to the absence of microrotation velocity field. Ramkissoon & Majumdar [11] linearized the governing equations of micropolar fluids and applied Fourier transforms to obtain the three-dimensional micropolar Stokeslet. Olmstead & Majumdar [12] derived the two-dimensional micropolar Oseenlet and micropolar Oseen couplet. In this paper, we derive fundamental Stokes and Oseen solutions of micropolar flows in three dimensions, so that the point force and point couple can be prescribed in any direction. Corresponding results for two-dimensional flows are also presented.

## 2. Stokes and Oseen flows of a micropolar fluid due to a point force

Consider a point force in an unbounded, quiescent, incompressible micropolar fluid. Without loss of generality, the point force is placed at the origin, and the free-stream velocity $\mathbf{U}_\infty$ is taken to be $(U_\infty, 0, 0)$. The resultant fluid flow is assumed steady. Based on the Oseen approximation, the governing equations [13] reduce to

$$\nabla \cdot \mathbf{u} = 0, \tag{1}$$

$$\rho U_\infty \frac{\partial \mathbf{u}}{\partial x_1} = -\nabla p - (\mu + \mu_r)\nabla \times \nabla \times \mathbf{u} + 2\mu_r \nabla \times \mathbf{v} + \rho \mathbf{F}\delta(\mathbf{x}), \tag{2}$$

$$\rho I_r U_\infty \frac{\partial \mathbf{v}}{\partial x_1} = 2\mu_r(\nabla \times \mathbf{u} - 2\mathbf{v}) - c_r \nabla \times \nabla \times \mathbf{v} + c_m \nabla(\nabla \cdot \mathbf{v}), \tag{3}$$

where $\rho$ is the fluid density, $I_r$ is the microinertia, $\mu$, $\mu_r$, $c_r$ and $c_m$ are the Newtonian, microrotational, and two angular viscosities, respectively, and $\mathbf{F}$ is a constant vector. In the equation (3), the divergence of $\mathbf{v}$ is assumed zero, which is verified in the latter part of this section. The pressure, $p$, translational velocity, $\mathbf{u}$, and microrotation velocity, $\mathbf{v}$, are required to decay as $|\mathbf{x}| \to \infty$ in an unbounded flow,

$$p \to 0, \ |\mathbf{u}| \to 0, \ |\mathbf{v}| \to 0 \text{ as } |\mathbf{x}| \to \infty.$$

Suppose that $f$ is an absolutely integrable function that decays at infinity of $\mathfrak{R}^n$. The $n$-dimensional complex Fourier transform of the function $f$, is defined by

$$\hat{f}(\boldsymbol{\xi}) = \mathcal{F}\{f(\mathbf{x})\} = (2\pi)^{-\frac{n}{2}} \int_{\mathfrak{R}^n} f(\mathbf{x}) e^{-i\boldsymbol{\xi} \cdot \mathbf{x}} d\mathbf{x},$$

where $\boldsymbol{\xi}$ is the transformed variable of $\mathbf{x}$, $\boldsymbol{\xi} \cdot \mathbf{x} = \xi_1 x_1 + \cdots + \xi_n x_n$, and $i$ is the imaginary unit, $i = \sqrt{-1}$.

The divergence of (2) yields

$$\nabla^2 p = \rho \nabla \cdot [\mathbf{F}\delta(\mathbf{x})], \tag{4}$$

which states that $p$ is harmonic everywhere except at the pole. To solve (4) for $p$, we take the Fourier transform, finding

$$\hat{p} = -\rho(i\boldsymbol{\xi}) \cdot \mathbf{F}\left[\frac{(2\pi)^{-\frac{n}{2}}}{\xi^2}\right].$$

Taking the inverse Fourier transform, we find

$$p = \begin{cases} \rho \nabla \cdot \left(\mathbf{F}\dfrac{1}{2\pi}\ln r\right) = \dfrac{\rho}{2\pi}\dfrac{\mathbf{F}\cdot\mathbf{x}}{r^2} & n = 2, \\ \rho \nabla \cdot \left(-\mathbf{F}\dfrac{1}{4\pi r}\right) = \dfrac{\rho}{4\pi}\dfrac{\mathbf{F}\cdot\mathbf{x}}{r^3} & n = 3. \end{cases} \tag{5}$$

Stokes and Oseen flows due to a point force have the same pressure field, regardless of whether the fluid is Newtonian or micropolar.

From (3), we have

$$\nabla \times \mathbf{u} = 2\mathbf{v} + \frac{c_r}{2\mu_r}\nabla \times \nabla \times \mathbf{v} + \frac{\rho I_r}{2\mu_r}(\mathbf{U}_\infty \cdot \nabla)\mathbf{v}. \tag{6}$$

Taking the curl of (2) gives

$$-(\mu + \mu_r)\nabla \times \nabla \times \nabla \times \mathbf{u} + 2\mu_r \nabla \times \nabla \times \mathbf{v} + \rho \nabla \times [\mathbf{F}\delta(\mathbf{x})] = \rho(\mathbf{U}_\infty \cdot \nabla)(\nabla \times \mathbf{u}). \tag{7}$$

Substituting (6) in (7), we derive a partial differential vector equation containing only one unknown, $\mathbf{v}$,

$$\left(\nabla^4 - \lambda^2 \nabla^2 - a_1 \nabla^2 \frac{\partial}{\partial x_1} + a_2 \frac{\partial}{\partial x_1} + a_3 \frac{\partial^2}{\partial x_1^2}\right)\mathbf{v} = a_4 \nabla \times [\mathbf{F}\delta(\mathbf{x})], \tag{8}$$

where $\lambda^2 = \dfrac{4\mu_r \mu}{c_r(\mu + \mu_r)}$, $a_1 = \rho U_\infty \left(\dfrac{I_r}{c_r} + \dfrac{1}{\mu + \mu_r}\right)$, $a_2 = \dfrac{4\rho U_\infty \mu_r}{c_r(\mu + \mu_r)}$, $a_3 = \dfrac{\rho^2 U_\infty^2 I_r}{c_r(\mu + \mu_r)}$, and

$a_4 = \dfrac{2\rho \mu_r}{c_r(\mu + \mu_r)}$.

To solve partial differential equations of high order, such as (8), we may factorize the high order partial differential operator into products of lower order [14]. This method was used by Olmstead & Majumdar [12]. Formally, it is proposed that

$$L = \left(\nabla^2 + A_1 \frac{\partial}{\partial x_1} + B_1\right)\left(\nabla^2 + A_2 \frac{\partial}{\partial x_1} + B_2\right), \tag{9}$$

where $L$ is a fourth order partial differential operator, and $A_1$, $A_2$, $B_1$ and $B_2$ are constants. While the method of factorization is attractive, a certain relationship between the parameters must exist for $L$ to admit the desired factorization. To factorize the differential operator in (8), the following must be true:

$$\begin{aligned} L &= \nabla^4 - \lambda^2 \nabla^2 - a_1 \nabla^2 \frac{\partial}{\partial x_1} + a_2 \frac{\partial}{\partial x_1} + a_3 \frac{\partial^2}{\partial x_1^2} \\ &= \nabla^4 + (B_1 + B_2)\nabla^2 + (A_1 + A_2)\nabla^2 \frac{\partial}{\partial x_1} + (A_1 B_2 + B_1 A_2)\frac{\partial}{\partial x_1} + A_1 A_2 \frac{\partial^2}{\partial x_1^2} + B_1 B_2. \end{aligned} \tag{10}$$

Consequently, it is required that

$$B_1 + B_2 = -\lambda^2, \; A_1 + A_2 = -a_1, \; A_1 B_2 + B_1 A_2 = a_2, \; A_1 A_2 = a_3, \; B_1 B_2 = 0.$$

We have five equations and only four unknowns. To expedite the solution, the value of $B_1$ is taken to be zero since $B_1 B_2 = 0$. Therefore,

$$B_1 = 0 \Rightarrow B_2 = -\lambda^2.$$

Then, from $A_1 B_2 + B_1 A_2 = a_2$,

$$A_1 = -\frac{a_2}{\lambda^2} = -\frac{\rho U_\infty}{\mu}.$$

Consequently,

$$A_2 = \frac{a_3}{A_1} = -\frac{\rho\mu U_\infty I_r}{c_r(\mu+\mu_r)}.$$

However, $A_1 + A_2 = -a_1$, or

$$\frac{\rho U_\infty}{\mu} + \frac{\rho\mu U_\infty I_r}{c_r(\mu+\mu_r)} = \rho U_\infty\left(\frac{I_r}{c_r} + \frac{1}{\mu+\mu_r}\right),$$

which gives

$$I_r = \frac{c_r}{\mu}. \tag{11}$$

Hence, the partial differential operator in (10) can be factorized into

$$\nabla^4 - \lambda^2\nabla^2 - a_1\nabla^2\frac{\partial}{\partial x_1} + a_2\frac{\partial}{\partial x_1} + a_3\frac{\partial^2}{\partial x_1^2} = \left(\nabla^2 - \frac{a_2}{\lambda^2}\frac{\partial}{\partial x_1}\right)\left(\nabla^2 - \frac{a_3\lambda^2}{a_2}\frac{\partial}{\partial x_1} - \lambda^2\right).$$

This allows (8) to be rewritten as

$$\left(\nabla^2 - 2n_0\frac{\partial}{\partial x_1}\right)\left(\nabla^2 - 2m_0\frac{\partial}{\partial x_1} - \lambda^2\right)\mathbf{v} = a_4\nabla\times[\mathbf{F}\delta(\mathbf{x})], \tag{12}$$

where $2n_0 = \frac{\rho U_\infty}{\mu}$ and $2m_0 = \frac{\rho U_\infty}{\mu+\mu_r}$. The above factorization is valid under the physical constraint of the parameters given by (11).

To solve (12) for $\mathbf{v}$, it is convenient to take the Fourier transform.

$$\left(\xi^2 + i2n_0\xi_1\right)\left(\xi^2 + i2m_0\xi_1 + \lambda^2\right)\hat{\mathbf{v}} = (2\pi)^{-\frac{n}{2}}a_4(i\boldsymbol{\xi})\times\mathbf{F},$$

which gives

$$\hat{\mathbf{v}} = \frac{(2\pi)^{-\frac{n}{2}}a_4(i\boldsymbol{\xi})\times\mathbf{F}}{\left(\xi^2 + i2n_0\xi_1\right)\left(\xi^2 + i2m_0\xi_1 + \lambda^2\right)}. \tag{13}$$

The inverse Fourier transform gives

$$\mathbf{v} = a_4\nabla\times\left\{\mathbf{F}\mathcal{F}^{-1}\left[\frac{(2\pi)^{-\frac{n}{2}}}{\left(\xi^2 + i2n_0\xi_1\right)\left(\xi^2 + i2m_0\xi_1 + \lambda^2\right)}\right]\right\},$$

which is the micropolar Oseenlet of $\mathbf{v}$ (see Appendix)

$$\mathbf{v} = \begin{cases} \dfrac{2\mu}{c_r} \nabla \times \left\{ \mathbf{F} \int_{x_1}^{\infty} e^{\alpha_0(x_1-t)} \dfrac{e^{n_0 t} K_0(n_0 s) - e^{m_0 t} K_0(w_0 s)}{2\pi U_\infty} dt \right\} & n = 2, \\ \dfrac{2\mu}{c_r} \nabla \times \left[ \mathbf{F} \int_{x_1}^{\infty} e^{\alpha_0(x_1-t)} \dfrac{e^{n_0(t-s)} - e^{m_0 t - w_0 s}}{4\pi U_\infty s} dt \right] & n = 3, \end{cases} \qquad (14)$$

where $\alpha_0 = \dfrac{\lambda^2}{2(n_0 - m_0)} = \dfrac{4\mu^2}{\rho U_\infty c_r}$, $w_0 = \sqrt{m_0^2 + \lambda^2}$, $s = \sqrt{t^2 + \sum_{j=2}^{n} x_j^2}$, and $K_0(\xi) = \int_0^\infty e^{-\xi \cosh \tau} d\tau$ is the modified Bessel function of the second kind. Because $\mathbf{v}$ is expressed as the curl of the product of a scalar function and the constant vector $\mathbf{F}$, the divergence of $\mathbf{v}$ is zero. Hence, the assumption made earlier about the divergence of $\mathbf{v}$ being zero is satisfied.

As $U_\infty \to 0$, (14) produces the micropolar Stokeslet,

$$\mathbf{v} = \begin{cases} -\dfrac{\rho}{4\pi\mu} \nabla \times \{ \mathbf{F}[\ln r + K_0(\lambda r)] \} & n = 2, \\ \dfrac{\rho}{8\pi\mu} \nabla \times \left( \mathbf{F} \dfrac{1 - e^{-\lambda r}}{r} \right) & n = 3. \end{cases} \qquad (15)$$

Equation (15) gives the microrotation velocity in the presence of a point force.

The curl of (3) gives the curl of the curl of $\mathbf{u}$ as

$$\nabla \times \nabla \times \mathbf{u} = 2\nabla \times \mathbf{v} + \dfrac{c_r}{2\mu_r} \nabla \times \nabla \times \nabla \times \mathbf{v} + \dfrac{\rho I_r U_\infty}{2\mu_r} \dfrac{\partial}{\partial x_1} (\nabla \times \mathbf{v}),$$

which can be written as

$$-\nabla^2 \mathbf{u} = \left( 2 - \dfrac{c_r}{2\mu_r} \nabla^2 + \dfrac{\rho I_r U_\infty}{2\mu_r} \dfrac{\partial}{\partial x_1} \right) (\nabla \times \mathbf{v}),$$

using vector identities, (1) and the assumption that the divergence of $\mathbf{v}$ is zero. Taking the Fourier transform, we find

$$\xi^2 \hat{\mathbf{u}} = \left( 2 + \dfrac{c_r}{2\mu_r} \xi^2 + i \dfrac{\rho I_r U_\infty}{2\mu_r} \xi_1 \right) (i\boldsymbol{\xi}) \times \hat{\mathbf{v}}.$$

Substituting (11) and (13) in the above equation leads to

$$\hat{\mathbf{u}} = (2\pi)^{-\frac{n}{2}} a_4 (i\boldsymbol{\xi}) \times (i\boldsymbol{\xi}) \times \left\{ \mathbf{F} \left[ \dfrac{2}{\xi^2 (\xi^2 + i2n_0 \xi_1)(\xi^2 + i2m_0 \xi_1 + \lambda^2)} + \dfrac{c_r}{2\mu_r} \dfrac{1}{\xi^2 (\xi^2 + i2m_0 \xi_1 + \lambda^2)} \right] \right\}.$$

The inverse Fourier transform of this expression yields $\mathbf{u}$ in the form

$$\mathbf{u} = a_4 \nabla \times \nabla \times \left\{ 2\mathbf{F} \mathcal{F}^{-1} \left[ \dfrac{(2\pi)^{-\frac{n}{2}}}{\xi^2 (\xi^2 + i2n_0 \xi_1)(\xi^2 + i2m_0 \xi_1 + \lambda^2)} \right] + \dfrac{c_r}{2\mu_r} \mathbf{F} \mathcal{F}^{-1} \left[ \dfrac{(2\pi)^{-\frac{n}{2}}}{\xi^2 (\xi^2 + i2m_0 \xi_1 + \lambda^2)} \right] \right\}.$$

Finally, we find the translational velocity **u** is given by the micropolar Oseenlet of **u** (see Appendix)

$$\mathbf{u} = \begin{cases} \nabla \times \nabla \times \left\{ \mathbf{F} \int_{x_1}^{\infty} \dfrac{\left[1 - e^{\alpha_0(x_1 - t)}\right] e^{n_0 t} K_0(n_0 s) + e^{\alpha_0(x_1 - t)} e^{m_0 t} K_0(w_0 s) + \ln s}{2\pi U_\infty} dt \right\} & n = 2, \\ \nabla \times \nabla \times \left\{ \mathbf{F} \int_{x_1}^{\infty} \dfrac{\left[1 - e^{\alpha_0(x_1 - t)}\right] e^{n_0(t-s)} + e^{\alpha_0(x_1 - t)} e^{m_0 t - w_0 s} - 1}{4\pi U_\infty s} dt \right\} & n = 3. \end{cases} \quad (16)$$

In the limit $\mu_r \to 0$, the translational velocity and microrotation velocity fields decouple. Then, $m_0 \to n_0$, $w_0 \to n_0$, and the expression (16) of **u** simplifies to

$$\mathbf{u} = \begin{cases} \nabla \times \nabla \times \left[ \mathbf{F} \int_{x_1}^{\infty} \dfrac{e^{n_0 t} K_0(n_0 s) + \ln s}{2\pi U_\infty} dt \right] & n = 2, \\ \nabla \times \nabla \times \left[ \mathbf{F} \int_{x_1}^{\infty} \dfrac{e^{n_0(t-s)} - 1}{4\pi U_\infty s} dt \right] & n = 3. \end{cases} \quad (17)$$

We see that the Newtonian Oseenlet is recovered.

In the limit $U_\infty \to 0$, the micropolar Oseenlet of **u** in (16) becomes the micropolar Stokeslet of **u**,

$$\mathbf{u} = \begin{cases} \nabla \times \nabla \times \left\{ \mathbf{F} \left[ \dfrac{\rho r^2 \ln r}{8\pi\mu} + \dfrac{\rho c_r}{8\mu^2 \pi} (\ln r + K_0(\lambda r)) \right] \right\} & n = 2, \\ -\nabla \times \nabla \times \left\{ \mathbf{F} \left[ \dfrac{\rho r}{8\pi\mu} + \dfrac{\rho c_r}{16\mu^2 \pi} \left( \dfrac{1 - e^{-\lambda r}}{r} \right) \right] \right\} & n = 3. \end{cases} \quad (18)$$

The solution of **u** for a micropolar fluid is much more complicated than that for a Newtonian fluid.

## 3. Stokes and Oseen flows of a micropolar fluid due to a point couple

Consider a point couple in an unbounded quiescent, incompressible micropolar fluid. Based on the Oseen approximation, the governing equations [13] can be linearized as

$$\nabla \cdot \mathbf{u} = 0, \quad (19)$$

$$\rho U_\infty \dfrac{\partial \mathbf{u}}{\partial x_1} = -\nabla p - (\mu + \mu_r) \nabla \times \nabla \times \mathbf{u} + 2\mu_r \nabla \times \mathbf{v}, \quad (20)$$

$$\rho I_r U_\infty \dfrac{\partial \mathbf{v}}{\partial x_1} = 2\mu_r (\nabla \times \mathbf{u} - 2\mathbf{v}) - c_r \nabla \times \nabla \times \mathbf{v} + c_m \nabla(\nabla \cdot \mathbf{v}) + \rho \mathbf{T} \delta(\mathbf{x}), \quad (21)$$

where $\rho \mathbf{T} \delta(\mathbf{x})$ is the point couple, with $\mathbf{T}$ as a constant vector. Without loss of generality, the point couple is assumed to be positioned at the origin. We begin by taking the divergence of (20), which states

$$\nabla^2 p = 0.$$

Because $p \to 0$ as $|\mathbf{x}| \to \infty$, the pressure field $p$ is such that

$$p = 0.$$

This reduces the gradient of $p$ in (20) to zero.

To obtain the translational velocity field $\mathbf{u}$, we take the curl of (21),

$$c_r \nabla^2 \mathbf{a} - 2\mu_r \nabla^2 \mathbf{u} - 4\mu_r \mathbf{a} + \rho \nabla \times [\mathbf{T}\delta(\mathbf{x})] = \rho I_r U_\infty \frac{\partial \mathbf{a}}{\partial x_1}, \tag{22}$$

where $\mathbf{a} = \nabla \times \mathbf{v}$. Vector identities and (19) were used to express the curl of (21) in the above form. To express (22) in terms of $\mathbf{u}$ alone, we make use of (20), which can be rewritten as

$$\mathbf{a} = \frac{1}{2\mu_r} \left[ \rho U_\infty \frac{\partial}{\partial x_1} - (\mu + \mu_r) \nabla^2 \right] \mathbf{u}. \tag{23}$$

Substituting (23) in (22) leads to

$$\left( \nabla^4 - \lambda^2 \nabla^2 - a_1 \frac{\partial}{\partial x_1} \nabla^2 + a_2 \frac{\partial}{\partial x_1} + a_3 \frac{\partial^2}{\partial x_1^2} \right) \mathbf{u} = a_4 \nabla \times [\mathbf{T}\delta(\mathbf{x})]. \tag{24}$$

Because (24) and (8) are identical in form, we can factorize the partial differential operator, as in (12), under the physical constraint given by (11). Then, we can write

$$\left( \nabla^2 - 2n_0 \frac{\partial}{\partial x_1} \right) \left( \nabla^2 - 2m_0 \frac{\partial}{\partial x_1} - \lambda^2 \right) \mathbf{u} = a_4 \nabla \times [\mathbf{T}\delta(\mathbf{x})],$$

whose Fourier transform is

$$\hat{\mathbf{u}} = \frac{(2\pi)^{-\frac{n}{2}} a_4 (i\boldsymbol{\xi}) \times \mathbf{T}}{\left( \xi^2 + i2n_0 \xi_1 \right) \left( \xi^2 + i2m_0 \xi_1 + \lambda^2 \right)}. \tag{25}$$

The inverse Fourier transform of $\hat{\mathbf{u}}$ yields the micropolar Oseen couplet of $\mathbf{u}$ (see Appendix)

$$\mathbf{u} = \begin{cases} \dfrac{2\mu}{c_r} \nabla \times \left\{ \mathbf{T} \displaystyle\int_{x_1}^{\infty} e^{\alpha_0 (x_1 - t)} \dfrac{e^{n_0 t} K_0(n_0 s) - e^{m_0 t} K_0(w_0 s)}{2\pi U_\infty} dt \right\} & n = 2, \\[2ex] \dfrac{2\mu}{c_r} \nabla \times \left[ \mathbf{T} \displaystyle\int_{x_1}^{\infty} e^{\alpha_0 (x_1 - t)} \dfrac{e^{n_0 (t-s)} - e^{m_0 t - w_0 s}}{4\pi U_\infty s} dt \right] & n = 3. \end{cases} \tag{26}$$

It is not surprising that the micropolar Oseen couplet of $\mathbf{u}$ in (26) is similar to the micropolar Oseenlet of $\mathbf{v}$ in (14), except that the former is caused by a point couple while the latter is due to a point force.

In the limit $U_\infty \to 0$, the micropolar Oseen couplet of $\mathbf{u}$ in (26) becomes the micropolar Stokes couplet of $\mathbf{u}$,

$$\mathbf{u} = \begin{cases} -\dfrac{\rho}{4\pi\mu} \nabla \times \{\mathbf{T}[\ln r + K_0(\lambda r)]\} & n = 2, \\ \dfrac{\rho}{8\pi\mu} \nabla \times \left(\mathbf{T}\dfrac{1-e^{-\lambda r}}{r}\right) & n = 3. \end{cases} \qquad (27)$$

We take the divergence of (21) to evaluate the divergence of $\mathbf{v}$ and find

$$\left(\nabla^2 - 2c_1 \frac{\partial}{\partial x_1} - \lambda_0^2\right) f = -c_2 \nabla \cdot [\mathbf{T}\delta(\mathbf{x})], \qquad (28)$$

where $f = \nabla \cdot \mathbf{v}$, $2c_1 = \dfrac{\rho I_r U_\infty}{c_m}$, $\lambda_0^2 = \dfrac{4\mu_r}{c_m}$ and $c_2 = \dfrac{\rho}{c_m}$. To expedite the solution, we take the Fourier transform of (28), and find

$$\hat{f} = \frac{(2\pi)^{-\frac{n}{2}} c_2 (i\boldsymbol{\xi}) \cdot \mathbf{T}}{\xi^2 + i2c_1\xi_1 + \lambda_0^2}, \qquad (29)$$

whose inverse Fourier transform is

$$f = \begin{cases} c_2 \nabla \cdot \left[\mathbf{T}\dfrac{e^{c_1 x_1} K_0(w_1 r)}{2\pi}\right] & n = 2, \\ c_2 \nabla \cdot \left[\mathbf{T}\dfrac{e^{(c_1 x_1 - w_1 r)}}{4\pi r}\right] & n = 3, \end{cases}$$

where $w_1 = \sqrt{c_1^2 + \lambda_0^2}$.

To determine the curl of $\mathbf{v}$, we take the Fourier transform of (23) and substitute (25) to find

$$\hat{\mathbf{a}} = \frac{a_4(\mu+\mu_r)}{2\mu_r} \frac{(2\pi)^{-\frac{n}{2}}(\xi^2 + i2m_0\xi_1)(i\boldsymbol{\xi})\times\mathbf{T}}{(\xi^2 + i2n_0\xi_1)(\xi^2 + i2m_0\xi_1 + \lambda^2)}. \qquad (30)$$

To relate $f$ to $\mathbf{a}$, we make use of the vector identity:

$$\nabla^2 \mathbf{v} = \nabla f - \nabla \times \mathbf{a},$$

whose Fourier transform is

$$-\xi^2 \hat{\mathbf{v}} = (i\boldsymbol{\xi})\hat{f} - (i\boldsymbol{\xi})\times\hat{\mathbf{a}}.$$

We substitute (29) and (30) in the above equation to express $\hat{\mathbf{v}}$ in terms of variables $(\xi_1, \xi_2, \xi_3)$,

$$\hat{\mathbf{v}} = c_2 (i\xi)(i\xi) \cdot \left[ \frac{-(2\pi)^{-\frac{n}{2}} \mathbf{T}}{\xi^2 (\xi^2 + i2c_1\xi_1 + \lambda_0^2)} \right]$$
$$+ \frac{a_4(\mu + \mu_r)}{2\mu_r} (i\xi) \times (i\xi) \times \left[ \frac{(2\pi)^{-\frac{n}{2}} \mathbf{T}}{\xi^2 (\xi^2 + i2n_0\xi_1)} - \frac{\lambda^2 (2\pi)^{-\frac{n}{2}} \mathbf{T}}{\xi^2 (\xi^2 + i2n_0\xi_1)(\xi^2 + i2m_0\xi_1 + \lambda^2)} \right].$$

The inverse Fourier transform of $\hat{\mathbf{v}}$ is the micropolar Oseen couplet of $\mathbf{v}$ (see Appendix)

$$\mathbf{v} = \begin{cases} \frac{\mu}{c_r} \nabla \nabla \cdot \left[ \mathbf{T} \int_{-\infty}^{x_I} e^{-\beta_0(x_I - t)} \frac{e^{c_I t} K_0(w_I s) + \ln s}{2\pi U_\infty} dt \right] \\ -\frac{\mu}{c_r} \nabla \times \nabla \times \left\{ \mathbf{T} \int_{-\infty}^{x_I} e^{-\beta_0(x_I - t)} \frac{e^{m_0 t} K_0(w_0 s) + \ln s}{2\pi U_\infty} dt + \mathbf{T} \int_{x_I}^{\infty} e^{\alpha_0(x_I - t)} \frac{e^{m_0 t} K_0(w_0 s) - e^{n_0 t} K_0(n_0 s)}{2\pi U_\infty} dt \right\} & n = 2, \\ \frac{\mu}{c_r} \nabla \nabla \cdot \left[ \mathbf{T} \int_{-\infty}^{x_I} e^{-\beta_0(x_I - t)} \frac{e^{c_I t - w_I s} - 1}{4\pi U_\infty s} dt \right] \\ -\frac{\mu}{c_r} \nabla \times \nabla \times \left\{ \mathbf{T} \int_{-\infty}^{x_I} e^{-\beta_0(x_I - t)} \frac{e^{m_0 t - w_0 s} - 1}{4\pi U_\infty s} dt + \mathbf{T} \int_{x_I}^{\infty} e^{\alpha_0(x_I - t)} \frac{e^{m_0 t - w_0 s} - e^{n_0(t-s)}}{4\pi U_\infty s} dt \right\} & n = 3, \end{cases}$$

(31)

where $\beta_0 = \frac{4\mu_r \mu}{\rho U_\infty c_r}$.

As $U_\infty \to 0$, (31) gives the micropolar Stokes couplet of $\mathbf{v}$

$$\mathbf{v} = \begin{cases} \frac{\rho}{8\pi\mu_r} \nabla \nabla \cdot \{\mathbf{T}[\ln r + K_0(\lambda_0 r)]\} - \frac{\rho(\mu + \mu_r)}{8\pi\mu_r \mu} \nabla \times \nabla \times \{\mathbf{T}[\ln r + K_0(\lambda r)]\} & n = 2, \\ -\frac{\rho}{16\pi\mu_r} \nabla \nabla \cdot \left( \mathbf{T} \frac{1 - e^{-\lambda_0 r}}{r} \right) + \frac{\rho(\mu + \mu_r)}{16\pi\mu_r \mu} \nabla \times \nabla \times \left( \mathbf{T} \frac{1 - e^{-\lambda r}}{r} \right) & n = 3. \end{cases}$$

(32)

The three-dimensional micropolar Stokes couplet of $\mathbf{v}$ almost agrees with that of Eringen [3], except for wrong sign in the first term.

## 4. Drag on a translating solid sphere in a micropolar viscous flow

Consider the flow produced by a solid sphere of radius $R$ translating with velocity $\mathbf{U}_\infty$ in an ambient micropolar fluid of infinite expanse. The flow due to the sphere may be obtained in terms of a point force and a potential dipole, both placed at the center of the sphere, as in the case of Stokes flow [15,16]. Hence, the velocity is given by

$$\mathbf{u} = -\nabla \times \nabla \times \left\{ \mathbf{F} \left[ \frac{\rho r}{8\pi\mu} + \frac{\rho c_r}{16\mu^2\pi} \left( \frac{1-e^{-\lambda r}}{r} \right) \right] \right\} + \mathbf{B} \bullet \nabla\nabla \frac{\rho}{4\pi r} + O(R_e) ,$$

where $\mathbf{B}$ is the vectorial strength of the potential dipole, and $R_e = \dfrac{\rho U_\infty 2R}{\mu}$ is the Reynolds number, assumed to be small. Requiring the boundary condition $\mathbf{u} = \mathbf{U}_\infty$ at $r = R$ on the surface of the sphere yields two algebraic equations for the coefficients $\mathbf{F}$ and $\mathbf{B}$,

$$\frac{\rho}{8\pi R(\mu+\mu_r)} \mathbf{F} - \frac{\rho}{4\pi R^3} \mathbf{B} = \mathbf{U}_\infty [1+O(R_e)] , \qquad \frac{\rho}{8\pi R^3(\mu+\mu_r)} \mathbf{F} + \frac{3\rho}{4\pi R^5} \mathbf{B} = \mathbf{O}[1+O(R_e)] ,$$

whose solution is

$$\rho \mathbf{F} = 6\pi R(\mu+\mu_r) \mathbf{U}_\infty [1+O(R_e)] , \qquad \rho \mathbf{B} = -\pi R^3 \mathbf{U}_\infty [1+O(R_e)] .$$

The drag comes exclusively from the point force. The dimensionless drag coefficient is

$$C_D = \frac{|\mathbf{F}|}{\frac{1}{2} U_\infty^2 \pi R^2} = \frac{24}{R_e} \frac{\mu+\mu_r}{\mu} [1+O(R_e)] .$$

Putting $\mu_r = 0$, we recover the result for the classical viscous flow

$$C_D = \frac{|\mathbf{F}|}{\frac{1}{2} U_\infty^2 \pi R^2} \approx \frac{24}{R_e} .$$

## 5. Drag on a translating circular cylinder in a micropolar viscous flow

Consider the flow produced by a circular cylinder of radius $R$ translating with velocity $\mathbf{U}_\infty$ in an ambient micropolar fluid of infinite expanse. The flow due to the cylinder can be obtained in terms of a two-dimensional point force and a two-dimensional potential dipole, both placed at the center of the cylinder,

$$\mathbf{u} = \nabla \times \nabla \times \left\{ \mathbf{F} \left[ \frac{\rho r^2 \ln r}{8\pi\mu} + \frac{\rho c_r}{8\mu^2\pi} (\ln r + K_0(\lambda r)) \right] \right\} + \mathbf{B} \bullet \nabla\nabla \left( -\frac{\rho}{2\pi} \ln r \right) + O(R_e) .$$

Requiring the boundary condition $\mathbf{u} = \mathbf{U}_\infty$ at $r = R$ on the surface of the cylinder, we find

$$\rho|\mathbf{F}| \approx \frac{8\pi(\mu+\mu_r) U_\infty}{1-2\gamma - 2\ln(R_e/8)}, \qquad \rho|\mathbf{B}| \approx -\frac{2\pi R^2 U_\infty}{1-2\gamma - 2\ln(R_e/8)} .$$

The drag comes exclusively from the two-dimensional point force. The dimensionless drag coefficient is

$$C_D = \frac{|\mathbf{F}|}{U_\infty^2 R} \approx \frac{16\pi}{R_e[1-2\gamma-2\ln(R_e/8)]} \frac{\mu+\mu_r}{\mu}.$$

By putting $\mu_r = 0$, the result for the classical two-dimensional viscous flow is recovered [17].

$$C_D = \frac{|\mathbf{F}|}{U_\infty^2 R} \approx \frac{16\pi}{R_e[1-2\gamma-2\ln(R_e/8)]}.$$

# 6. Conclusions

New fundamental solutions for micropolar fluids have been derived in explicit form. The problem of two- and three-dimensional, steady, unbounded Stokes and Oseen flows of a micropolar fluid due to a point force and a point couple was considered. The new fundamental solutions for Stokes and Oseen flows are the two-dimensional micropolar Stokeslet, given by (18) and (15), the two- and three-dimensional micropolar Stokes couplet, given by (27) and (32), the three-dimensional micropolar Oseenlet, given by (16) and (14), and the three-dimensional micropolar Oseen couplet, given by (26) and (31). These fundamental solutions are possible due to the existence of microrotation velocity fields in micropolar fluids. The fundamental solutions can generate further fundamental solutions by successive differentiation with respect to the singular point [15,16]. A summary of available fundamental solutions is given in Table 1.

Table 1: A list of new fundamental solutions derived in this paper

| singular point | dimensionality | Stokes flow | Oseen flow |
|---|---|---|---|
| A point force | 2-D | New | [12] |
| | 3-D | [11] | New |
| A point couple | 2-D | New | [12] |
| | 3-D | New | New |

These fundamental solutions for micropolar fluids can be used as the basic building blocks to construct new solutions of microscale flow problems by employing the boundary integral method or the singularity method. It was demonstrated that these fundamental solutions can be used to calculate the drag coefficients for a translating solid sphere and circular cylinder, respectively, in a micropolar fluid at low Reynolds numbers. The drag coefficients in a micropolar fluid are greater than those in a Newtonian fluid by the factor $\frac{\mu+\mu_r}{\mu}$.

# APPENDIX

Consider the partial differential equation

$$\left(\nabla^2 - 2b_1\frac{\partial}{\partial x_1} - b_2\right)\Phi_0 = -\delta(\mathbf{x}),$$

where $b_1$ and $b_2$ are two arbitrary constants. We apply the Fourier transform to obtain

$$\hat{\Phi}_0 = \frac{(2\pi)^{-\frac{n}{2}}}{\xi^2 + i2b_1\xi_1 + b_2}.$$

Letting $\Phi_0(\mathbf{x};b_1,b_2) = e^{b_1 x_1} g(r;b_1,b_2)$, we obtain

$$\left(\nabla^2 - 2b_1\frac{\partial}{\partial x_1} - b_2\right)\Phi_0 = e^{b_1 x_1}\left[\nabla^2 - \left(b_1^2 + b_2\right)\right]g,$$

and hence

$$\left[\nabla^2 - \left(b_1^2 + b_2\right)\right]g = -e^{-b_1 x_1}\delta(\mathbf{x}).$$

Taking the Fourier transform of the above equation, we have

$$\hat{g} = \frac{(2\pi)^{-\frac{n}{2}}}{\xi^2 + b_1^2 + b_2}.$$

The inverse Fourier transform of $\hat{g}$ is

$$g(r;b_1,b_2) = \begin{cases} \dfrac{K_0\left(r\sqrt{b_1^2 + b_2}\right)}{2\pi} & n = 2, \\ \dfrac{e^{-r\sqrt{b_1^2+b_2}}}{4\pi r} & n = 3. \end{cases}$$

Therefore, the fundamental solution $\Phi_0(\mathbf{x};b_1,b_2)$ is

$$\Phi_0(\mathbf{x};b_1,b_2) = \mathcal{F}^{-1}\left\{\frac{(2\pi)^{-\frac{n}{2}}}{\xi^2 + i2b_1\xi_1 + b_2}\right\} = \begin{cases} \dfrac{e^{b_1 x_1}K_0\left(r\sqrt{b_1^2 + b_2}\right)}{2\pi} & n = 2, \\ \dfrac{e^{b_1 x_1 - r\sqrt{b_1^2+b_2}}}{4\pi r} & n = 3. \end{cases}$$

In view of the properties of the Fourier transform, we find

$$\Phi(\mathbf{x};b_1,b_2,b_3) = \mathcal{F}^{-1}\left[\frac{(2\pi)^{-\frac{n}{2}}}{\left(\xi^2+i2b_1\xi_1+b_2\right)\left(i\xi_1+b_3\right)}\right] = \mathcal{F}^{-1}\left(\frac{\hat{\Phi}_0}{i\xi_1+b_3}\right) = -\int_{x_1}^{\infty} e^{-b_3(x_1-t)} e^{b_1 t} g(s;b_1,b_2)\,dt$$

$$\begin{cases} -\dfrac{1}{2\pi}\displaystyle\int_{x_1}^{\infty} e^{-b_3(x_1-t)} e^{b_1 t} K_0\!\left(s\sqrt{b_1^2+b_2}\right) dt & n=2, \\[2mm] -\dfrac{1}{4\pi}\displaystyle\int_{x_1}^{\infty} e^{-b_3(x_1-t)} \dfrac{e^{b_1 t - s\sqrt{b_1^2+b_2}}}{s}\,dt & n=3. \end{cases}$$

where $s = \sqrt{t^2 + \sum_{j=2}^{n} x_j^2}$ and $b_3$ is an arbitrary constant. To derive formulas (14), (16) (26) and (31), we perform the inverse Fourier transform:

$$\mathcal{F}^{-1}\left[\frac{(2\pi)^{-\frac{n}{2}}}{\xi^2\left(\xi^2+i2b_4\xi_1\right)\left(\xi^2+i2b_5\xi_1+b_6\right)}\right],$$

where $b_4$, $b_5$ and $b_6$ are three arbitrary constants. Partial fraction expansion gives

$$\frac{(2\pi)^{-\frac{n}{2}}}{\xi^2\left(\xi^2+i2b_4\xi_1\right)\left(\xi^2+i2b_5\xi_1+b_6\right)} = \frac{(2\pi)^{-\frac{n}{2}}}{b_4 b_6}\left[\frac{1}{i2\xi_1}\left(\frac{1}{\xi^2}-\frac{1}{\xi^2+i2b_4\xi_1}\right) - \frac{b_5}{i2b_5\xi_1+b_6}\left(\frac{1}{\xi^2}-\frac{1}{\xi^2+i2b_5\xi_1+b_6}\right) - \frac{b_4-b_5}{i2(b_5-b_4)\xi_1+b_6}\left(\frac{1}{\xi^2+i2b_4\xi_1}-\frac{1}{\xi^2+i2b_5\xi_1+b_6}\right)\right].$$

We conclude that

$$\mathcal{F}^{-1}\left[\frac{(2\pi)^{-\frac{n}{2}}}{\xi^2\left(\xi^2+i2b_4\xi_1\right)\left(\xi^2+i2b_5\xi_1+b_6\right)}\right] = \frac{1}{2b_4 b_6}\left[\Phi(\mathbf{x};0,0,0) - \Phi(\mathbf{x};b_4,0,0) - \Phi\!\left(\mathbf{x};0,0,\frac{b_6}{2b_5}\right) + \Phi\!\left(\mathbf{x};b_5,b_6,\frac{b_6}{2b_5}\right) \right.$$
$$\left. + \Phi\!\left(\mathbf{x};b_4,0,\frac{b_6}{2(b_5-b_4)}\right) - \Phi\!\left(\mathbf{x};b_5,b_6,\frac{b_6}{2(b_5-b_4)}\right)\right].$$